\documentclass[epj]{webofc}
\usepackage[utf8]{inputenc}
\usepackage[varg]{txfonts}   
\usepackage{booktabs}
\usepackage{xcolor}
\definecolor{darkred}{rgb}{0.4,0.0,0.0}
\definecolor{darkgreen}{rgb}{0.0,0.4,0.0}
\definecolor{darkblue}{rgb}{0.0,0.0,0.4}
\usepackage[bookmarks,linktocpage,colorlinks,
    linkcolor = darkred,
    urlcolor  = darkblue,
    citecolor = darkgreen]{hyperref}
\usepackage{subfigure}
\usepackage{slashed}
\wocname{EPJ Web of Conferences}
\woctitle{Lattice2017}
\newcommand{\tr}{{\rm Tr}}
\newcommand{\la}{\langle}
\newcommand{\ra}{\rangle}

\newcommand{\f}[2]{\frac{#1}{#2}}

\newcommand{\slaD}{{\slashed D}}

\begin{document}
%
\selectlanguage{english}
\title{%
Landau levels in QCD in an external magnetic field
}
\author{%
\firstname{Falk} \lastname{Bruckmann}\inst{1} \and
\firstname{Gergely} \lastname{Endr\H odi}\inst{2} \and
\firstname{Matteo}
\lastname{Giordano}\inst{3,4}\fnsep\thanks{Speaker,
  \email{giordano@bodri.elte.hu}} \and 
\firstname{S\'andor D.} \lastname{Katz}\inst{3,4} \and
\firstname{Tam\'as G.} \lastname{Kov\'acs}\inst{5} \and
\firstname{Ferenc} \lastname{Pittler}\inst{6} \and
\firstname{Jacob} \lastname{Wellnhofer}\inst{1}
}

\institute{%
Institute for Theoretical Physics, Universit\"at Regensburg, D-93040
Regensburg, Germany\and 
Institute for Theoretical Physics, Goethe Universit\"at Frankfurt,
D-60438 Frankfurt am Main, Germany\and 
E\"otv\"os University, Theoretical Physics, P\'azm\'any P.\ s.\ 1/A,
H-1117, Budapest, Hungary\and 
MTA-ELTE Lend\"ulet Lattice Gauge Theory Research Group, P\'azm\'any
P.\ s.\ 1/A, H-1117, Budapest, Hungary\and 
Institute of Nuclear Research of the Hungarian Academy of
Sciences, Bem t\'er 18/c, H-4026 Debrecen,  Hungary\and
HISKP(Theory), University of Bonn, Nussallee 14-16, D-53115 Bonn, Germany
}

\abstract{%
We will discuss the issue of Landau levels of quarks in lattice QCD in
an external magnetic field. We will show that in the two-dimensional
case the lowest Landau level can be identified unambiguously even if
the strong interactions are turned on. Starting from this observation,
we will then show how one can define a "lowest Landau level" in the
four-dimensional case, and discuss how much of the observed effects of
a magnetic field can be explained in terms of it. Our results can be
used to test the validity of low-energy models of QCD that make use of
the lowest-Landau-level approximation. 
}
\maketitle
\section{Introduction}\label{intro}

The effects of a background magnetic field on strongly interacting
matter are relevant to a number of problems in particle physics,
astrophysics and cosmology, see the recent
reviews~\cite{Kharzeev:2013jha,Miransky:2015ava}. In heavy ion
collisions, the colliding 
ions generate strong magnetic fields that affect the quark-gluon
plasma formed in the collision. Moreover, the equation of state of
hadronic matter in an external magnetic field is needed in the study 
of magnetars (a type of neutron stars), and also in the study of
the evolution of the early Universe. This provides enough motivation
to study the phase diagram of QCD in an external, uniform magnetic
field.

This task has been (partially) accomplished in recent years
by means of numerical calculations from first principles on the
lattice. Due to the yet unsolved sign problem at finite chemical
potential, only the corner of the phase diagram at vanishing chemical
potential can be reliably determined in this way. 
What is known in this case is
that for temperatures well below and well above the QCD pseudocritical
temperature $T_c$ at $B=0$, a constant magnetic field increases the quark
condensate, a phenomenon dubbed {\it magnetic catalysis} (MC), while around
$T_c$ it decreases the condensate, i.e., it gives {\it inverse magnetic
catalysis} (IMC)~\cite{Bali:2011qj}. This results in $T_c$ decreasing
with $B$. This behaviour was missed in early lattice
studies~\cite{DElia:2010abb}, due to the fact that lattice artefacts
can turn IMC into MC: fine lattices and physical quark
masses are needed to observe IMC on the lattice. 

An important question is how can these phenomena be understood in the
framework of QCD. To this end, it is convenient to distinguish the two
effects that an external magnetic field has on the formation of a
quark condensate. 
The magnetic field appears in fact both in the operator being averaged
(here the inverse of the Dirac operator) and in the fermionic
determinant, i.e., in the functional integral measure. The presence of
a magnetic field has been observed to 
increase the density of low modes of the Dirac operator in a typical
gauge configuration: this effect, dubbed {\it valence effect}, tends
to increase the condensate via the Banks-Casher relation, and
therefore pushes in the direction of MC. On the other hand,
configurations with a higher density of low modes are relatively
suppressed because of the determinant: this tends to reduce the
condensate and therefore  goes in the direction of IMC. This is the
so-called {\it sea effect}, which is typically weaker than the
valence effect, so that in general one finds 
MC. Near $T_c$, though, where the Polyakov loop effective potential is
flatter and so the sea effect is most effective, the 
latter ends up winning against the valence effect, therefore leading
to IMC~\cite{Bruckmann:2013oba}. 

IMC near $T_c$ was a quite unexpected result, given the pre-existing
common lore based on analytic calculations in low-energy models for
QCD, which predicted MC at all temperatures (see, e.g., the
reviews~\cite{Miransky:2015ava,Shovkovy:2012zn,Andersen:2014xxa}). The
basis for this common lore is the expected behaviour of the lowest
Landau level (LLL) of quarks in a constant magnetic field: as 
its degeneracy is proportional to the magnetic field, an increase in
the density of low Dirac modes is expected. This would explain the
valence effect, which is required for MC. In this framework, one
expects that only the LLL is physically relevant for sufficiently
large magnetic fields: this leads to the frequently used LLL
approximation, where all higher Landau levels are neglected (see,
e.g.,
Refs.~\cite{Leung:2005xz,Fayazbakhsh:2010gc,Fukushima:2011nu,Ferrer:2013noa}). 
However, in QCD quarks interact with the SU(3) gauge fields besides the
external magnetic field, and therefore the very meaning of Landau
levels is not obvious.
In the rest of this proceeding we will discuss in detail 
how the LLL can be interpreted in this setup,
how one can identify it on the lattice, and how much 
it actually
contributes to the (change of the) chiral condensate in a magnetic
field, so testing the validity of the LLL approximation. This work is
based on the paper~\cite{Bruckmann:2017pft}.

\section{Landau levels in the continuum}
\label{sec:LLcont}

As is well known, Landau levels (LLs) are the energy levels of an
otherwise 
free particle immersed in a uniform magnetic field $B$. For brevity,
in the following we will refer to this simply as the ``free case''.  
In the presence of interactions, one expects for weakly coupled
particles that the 
LL structure gets only slightly
perturbed. This has been verified on the lattice in the case of pions
in the presence of both strong interactions and an external magnetic
field~\cite{Bali:2011qj}. The  LLs
relevant to the problem of MC are however those of
quarks with ``Hamiltonian'' given by the Dirac operator in the
presence of non-Abelian gauge fields. Since quarks are strongly
coupled to the non-Abelian gauge fields, the question of whether the
LL structure survives the introduction of strong
interactions is a rather nontrivial one. 

We begin our discussion by reviewing the simplest possible case of
spin-$\f{1}{2}$ fermions of charge $q$ in the continuum, confined to
live in a two-dimensional plane, and subject only to a uniform
magnetic field orthogonal to that plane. In this case the eigenvalues
$\lambda_n^2$ of $-\slaD^2$ and their degeneracies $\nu_n$ read
\begin{equation}
  \label{eq:2dqcd}
 \lambda^2_{n}  = qB n\,, 
 \qquad   n = 2k + 1 -2s_z  \,,
 \qquad \nu_n = 
N_b (2-\delta_{0n})\,,
\end{equation}
where $k=0,1,\ldots$, $s_z=\pm\f{1}{2}$ is the spin of the particle
along the direction of the magnetic field, and so $n= 0,1,2,\ldots$,
and where $N_b = \f{L_x L_y qB}{2\pi}=\f{\Phi_B}{2\pi}\in \mathbb{Z}$
due to the quantisation of the magnetic flux $\Phi_B$ in a finite volume.
For quarks with $N_c$ colours one gets an extra 
factor $N_c$ in the
degeneracies. 
The generalisation to quarks in four dimensions is easy. The
eigenvalues get also contributions from the momenta in the $t$ and $z$
directions, while the degeneracies are doubled due to
particle-antiparticle symmetry: 
\begin{equation}
  \label{eq:4dqcd}
     \lambda^2_{n p_z p_t} = qB n 
+ p_z^2 + p_t^2\,,  \qquad
\nu_n = 
2 N_c N_b (2-\delta_{0n})\,.
\end{equation}
This result provides the basis for a simple argument for the valence
effect and in favour of MC.
If we believe that the 
LL structure is not affected very
much by the strong interactions, then since the LLL is just at 
$\lambda=0$ one would expect an increase in the spectral density near the
origin roughly proportional to $B$, and therefore an increase in the
chiral condensate. Assuming that the 
LL structure is
preserved is, however, quite a strong assumption that should be
carefully tested on the lattice, especially in the
light of the fact that quarks are strongly coupled. 


\section{Landau levels on the lattice: two-dimensional case}
\label{sec:LLlat2d}

\begin{figure}[t]
  \centering
\subfigure
{
  \includegraphics[width=0.45\textwidth,clip]{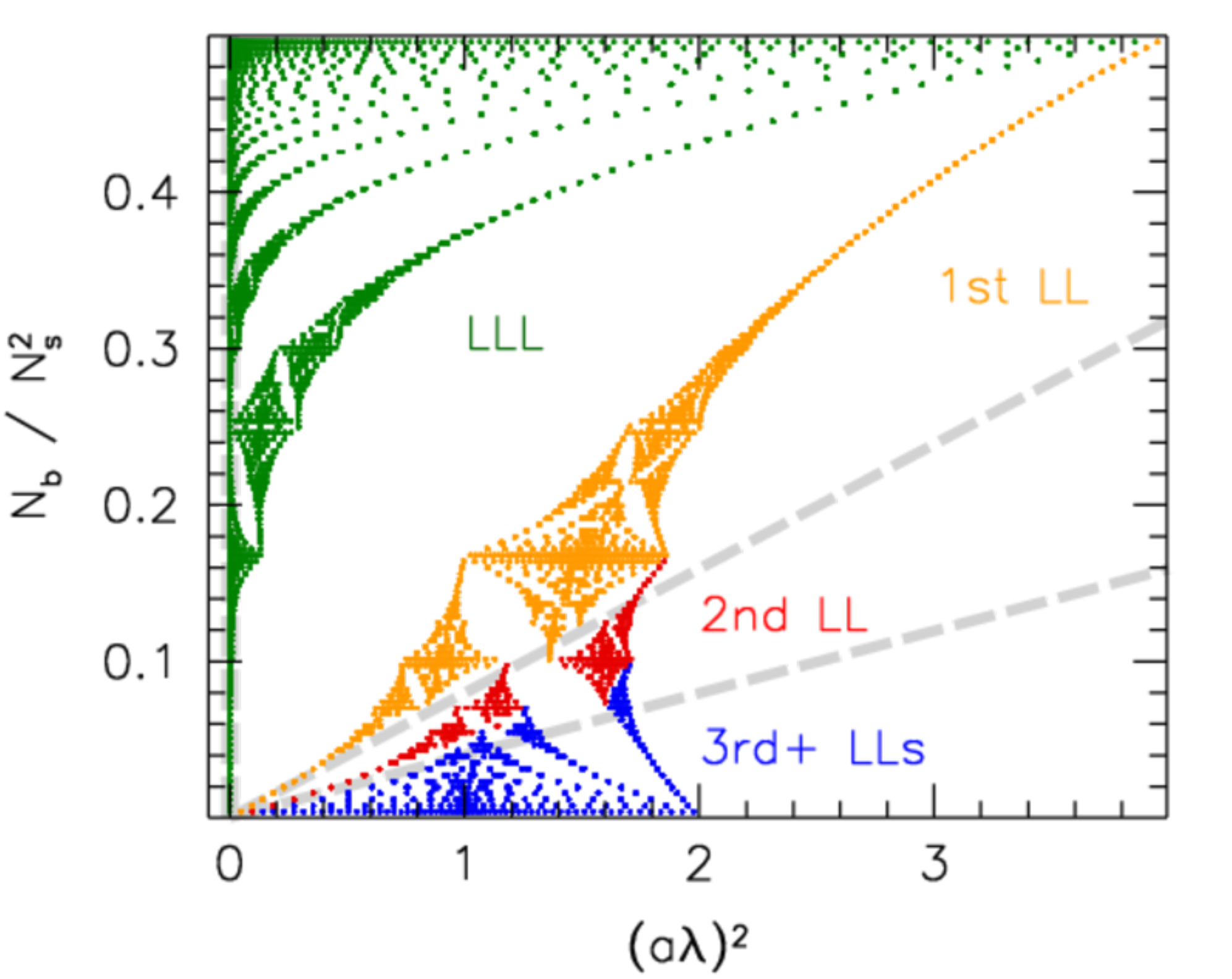}}
\subfigure{
  \includegraphics[width=0.45\textwidth,clip]{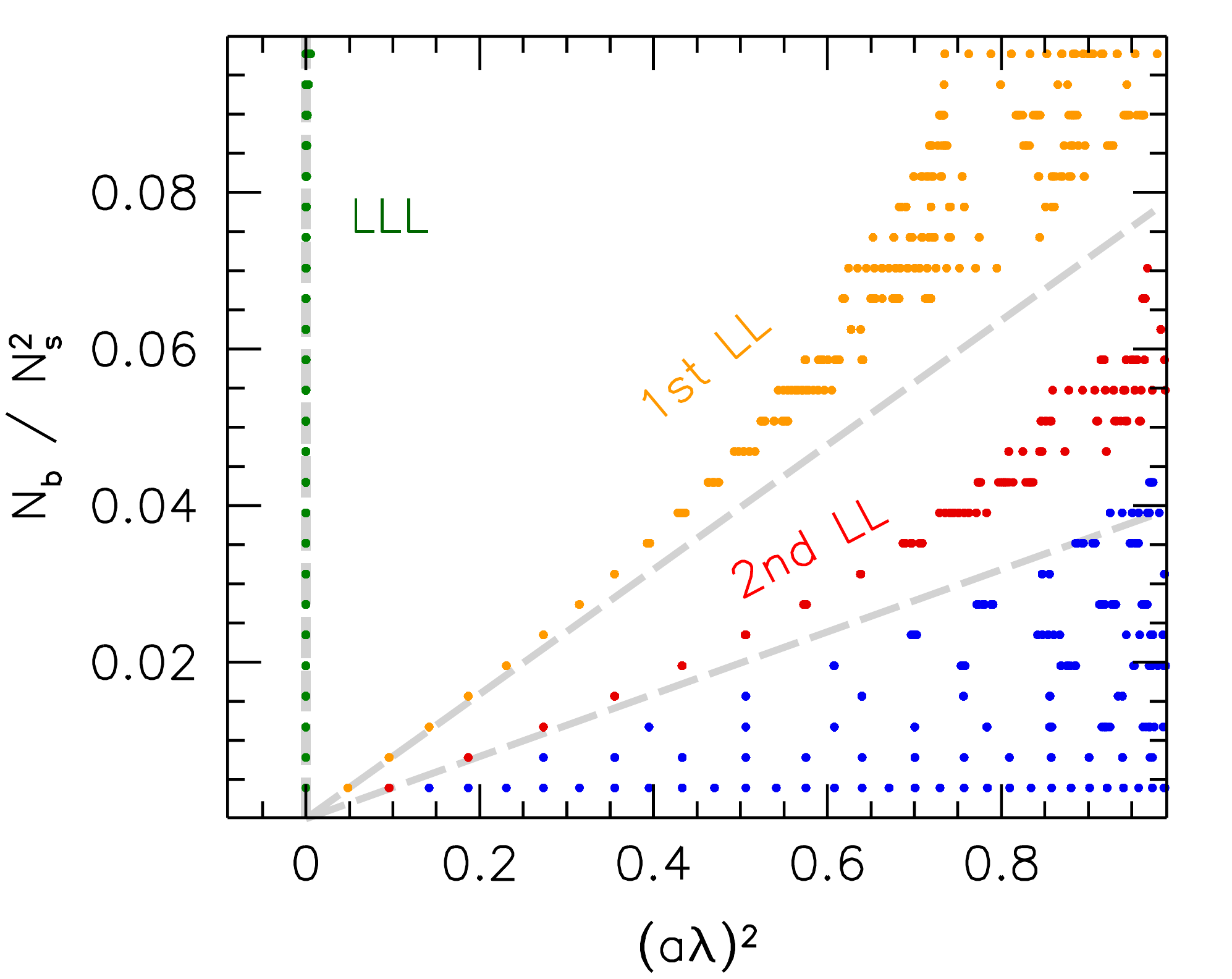}}
  \caption{Classification of the lattice eigenvalues according to continuum 
Landau level degeneracies in the free case. Continuum Landau levels
are shown as gray dashed lines. The right panel shows a zoom into the
region around the origin, where the continuum Landau levels are
approached.} 
  \label{fig:free}
\end{figure}

Our next step is to study the two-dimensional case on the lattice, using
the staggered Dirac operator $D_{\rm stag}$. This discretisation entails an exact
twofold doubling of the squared eigenvalues, which in the continuum
limit in the free case would lead to an extra factor of two in the
degeneracy of the LLs.
The results for the spectrum in the free case are well known (see
Fig.~\ref{fig:free}): the
eigenvalues spread around the continuum values due to lattice
artefacts, and give rise to a fractal structure known in the condensed
matter literature as Hofstadter's
butterfly~\cite{Hofstadter:1976zz,Endrodi:2014vza}. 
The LL structure is spoiled by the lattice, but nevertheless a remnant
of it still remains, as signalled by the presence of clear gaps
between groups of eigenvalues that match in size the degeneracies of
the continuum levels.

\begin{figure}[t]
  \centering
\subfigure{
  \includegraphics[width=0.45\textwidth,clip]{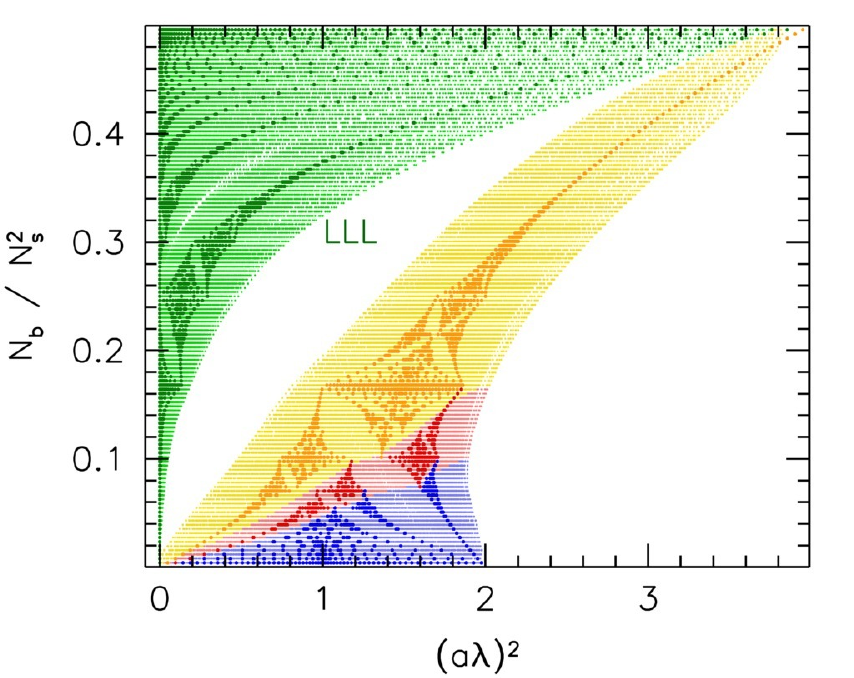}}
\subfigure{
  \includegraphics[width=0.45\textwidth,clip]{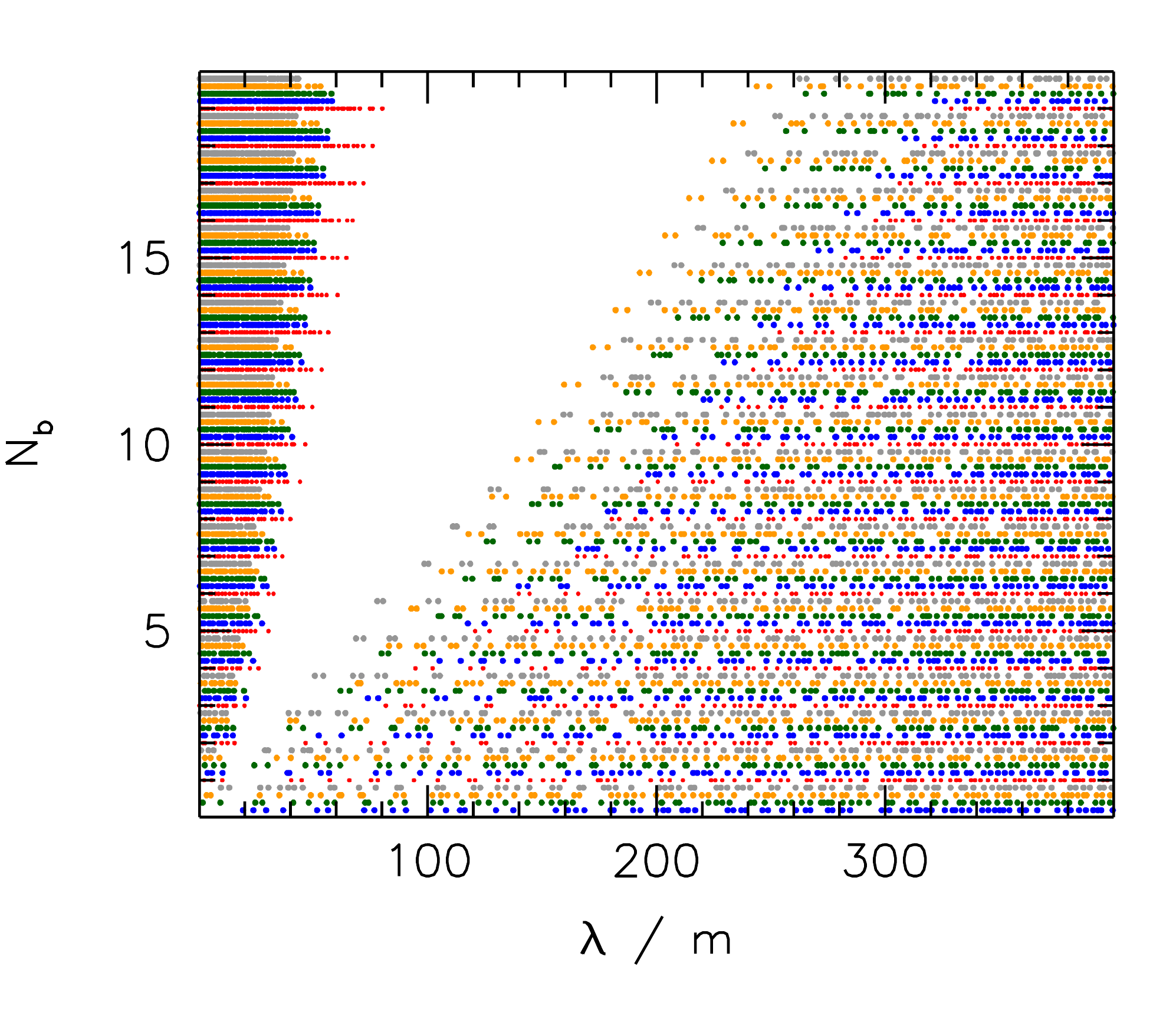}}
  \caption{Lattice eigenvalues in the interacting case. In the left
    panel eigenmodes from a single configuration are compared to the
    eigenmodes in the free case. The colour code corresponds 
to the classification into Landau levels according to the continuum
degeneracy. In the right panel, the gap between the LLL and the HLLs
in physical units is shown for various lattice spacings (corresponding
to the different colours) in the interacting case.} 
  \label{fig:int}
\end{figure}

We then switch on non-Abelian interactions, in practice by taking a 2D
spatial slice of a typical four-dimensional QCD configuration (for 2+1
dynamical flavours with physical masses at
$T\approx 400 \,{\rm  MeV}$). This results into the butterfly being
smeared out and the gaps disappearing, with a notable exception: the
lowest group of eigenvalues remains clearly separated from the rest
even in the presence of colour interactions (see
Fig.~\ref{fig:int}, left panel). Moreover, the number of 
eigenvalues in this group matches precisely those of the continuum
LLL. These features are not lattice artefacts: they
survive as the lattice is made finer and finer, if the appropriate
units are used, namely if eigenvalues are divided by the bare quark
mass (see Fig.~\ref{fig:int}, right panel). The notion of a LLL can
then be defined without ambiguity on the lattice also in the
interacting case. 
For brevity, higher modes will be collectively labelled as
higher Landau levels (HLLs), even though such higher levels do not
exist individually.

\subsection{Lowest Landau level and topology}
\label{sec:topor}

The reason behind the survival of the LLL in 2D in the presence of 
non-Abelian fields is its topological origin, which makes it robust
under (relatively) small deformations. In 2D the index theorem reads
$Q^{\rm 2D}_{\rm top} = N_\uparrow-N_\downarrow$, with $Q^{\rm
  2D}_{\rm top}$ the topological charge and
$N_{\uparrow},N_{\downarrow}$ the number of zero modes with  
spin-up and spin-down polarisation in the direction of the magnetic
field. Here 
the topological charge is just the magnetic flux  
(even in the presence of non-Abelian interactions),
\begin{equation}
  \label{eq:topch}
Q^{\rm 2D}_{\rm top} = \f{1}{2\pi} \int d^2 x\, F_{xy} = \f{1}{2\pi}\,
L^2 \cdot qB = N_b\,.
\end{equation}
Moreover, the ``vanishing
theorem''~\cite{Kiskis:1977vh,Nielsen:1977aw,Ansourian:1977qe} entails
that $N_\uparrow\cdot N_\downarrow=0$, and so for $qB>0$ 
there are exactly $N_b$ spin-up zero modes: these are precisely the
LLL modes in the continuum (for each colour). Notice
that these are the only modes with well-defined spin. When switching
on non-Abelian interactions, LLL modes survive as they are protected
by topology, while HLL modes mix and the corresponding structure is
washed away.

\begin{figure}[t]
 \centering
\subfigure{ \includegraphics[width=0.3\textwidth]{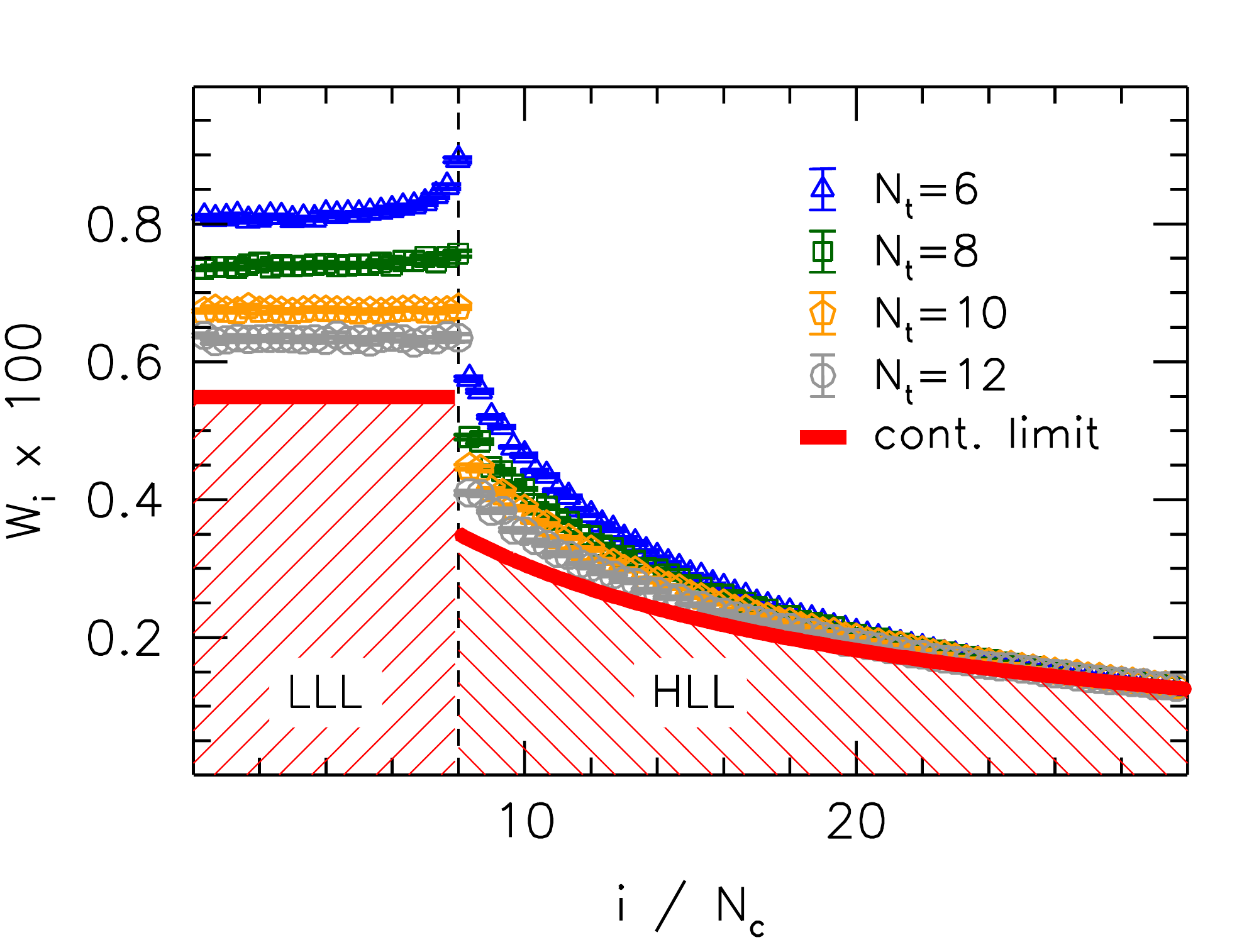}}
\subfigure{  \includegraphics[width=0.3\textwidth]{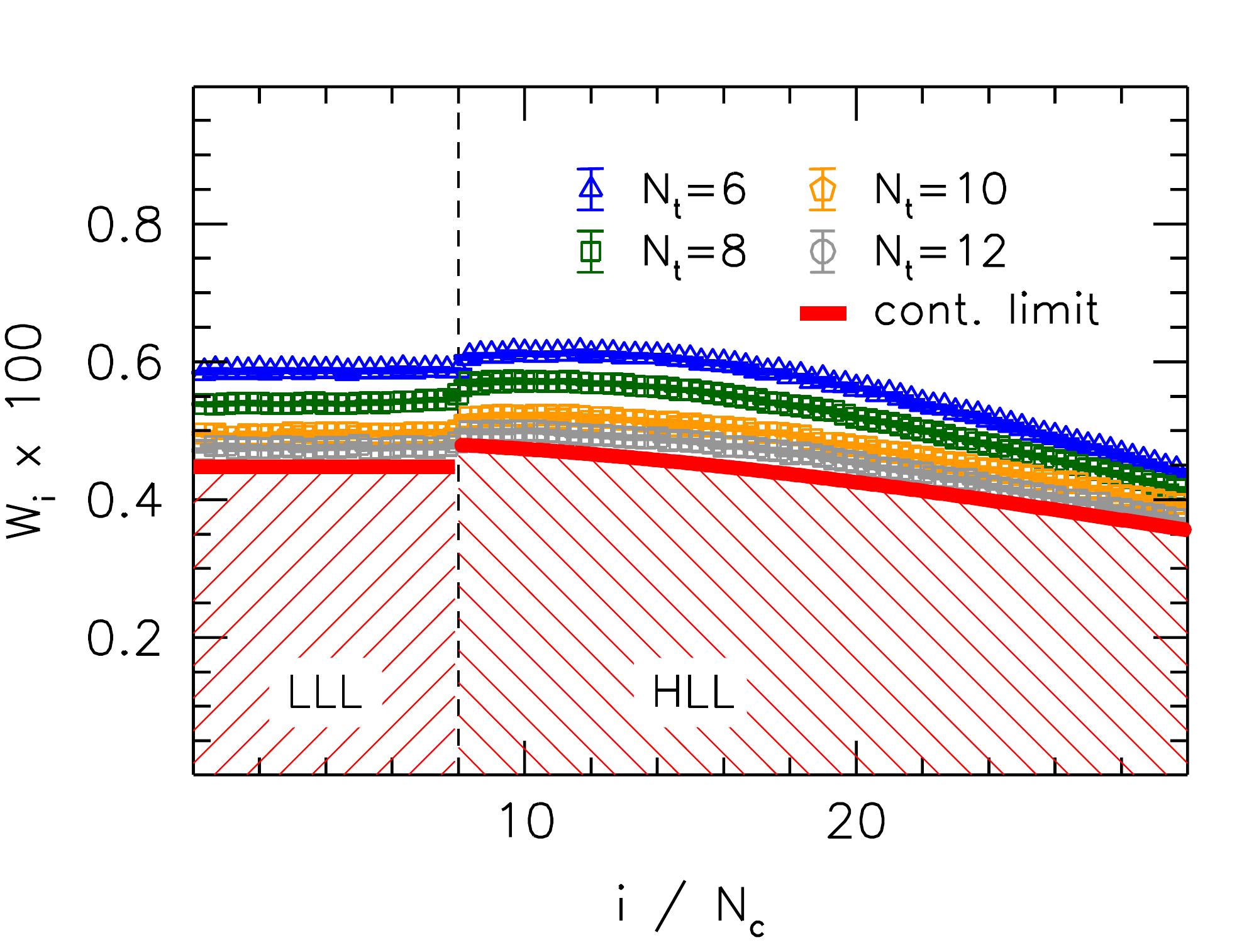}}
\subfigure{  \includegraphics[width=0.3\textwidth]{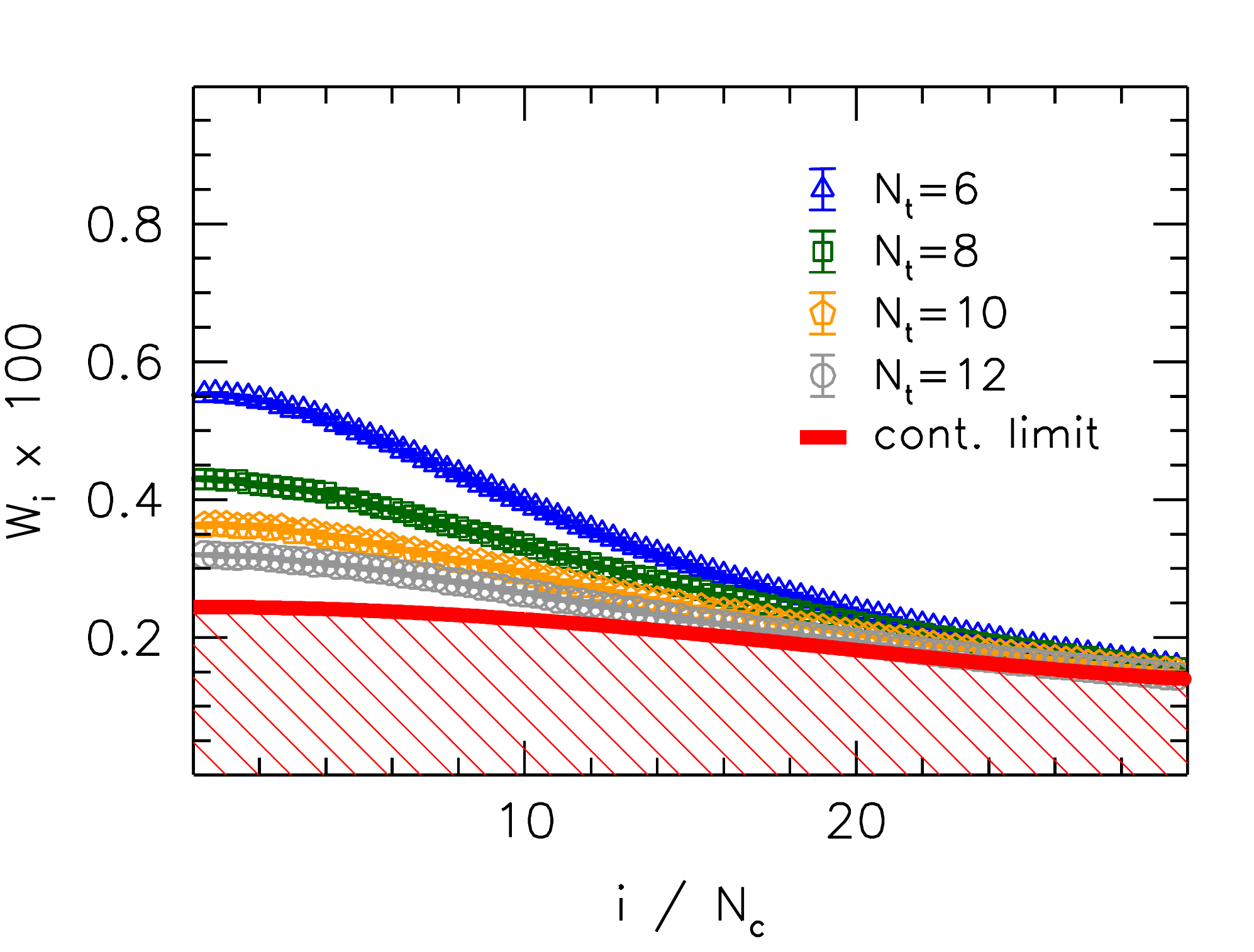}}
\caption{ The overlap Eq.~(\protect\ref{eq:overlapdef}) of 4D
 eigenmodes with the 2D modes as a function of the 2D index
 (in units of $N_c=3$) for a magnetic flux quantum
 $N_b=8$, on configurations generated at $T\approx400 \textmd{
   MeV}$. Left panel: low-lying 4D modes (with eigenvalue
 $220<\lambda/m<225$). Central panel: bulk 4D modes ($535<\lambda/m<545$). 
Right panel: same 4D modes as in the left panel, but overlap on 2D
modes at $N_b=0$.  
}
\label{fig:overlap}
\end{figure}

For staggered fermions 
the LLL modes are not exact
zero modes, but are expected to be (and indeed are) well separated
from the HLL eigenvalues. The identification of these near-zero modes
as LLL modes is further supported by the finding that the matrix
elements of the spin operator $\sigma_{xy} = \sigma_{z}$ are almost
perfectly diagonal, and considerably different from zero only for the
near-zero, LLL modes.

\section{Landau levels on the lattice: four-dimensional case}
\label{sec:LLlat4d}

In four dimensions the LLs 
cannot be identified 
by looking at the spectrum even in the free continuum case, as the
contribution of the $t$ and $z$ momenta make them overlap with each other. 
This is however not a major obstacle: the eigenmodes
$\tilde\psi^{(j)}_{ p_z  p_t}$ are in fact factorised, 
\begin{equation}
  \label{eq:modes}
  \tilde\psi^{(j)}_{ p_z  p_t}(x,y,z,t) = \phi^{(j)}(x,y)
  e^{ip_z z}e^{ip_t t} \,,
\end{equation}
with $\phi^{(j)}$ a solution of the 2D Dirac equation,
and so one can define a projector on the LLL (or on the
other levels, if so desired) and extract its contribution to any
observable:
\begin{equation}
  \label{eq:proj}
  \begin{aligned}
  P &= 
\sum_{j~\in~ {\rm LLL}}
\sum_{p_z, p_t} \tilde\psi^{(j)}_{ p_z
  p_t}\tilde\psi^{(j)\dag}_{ p_z p_t} = 
\sum_{j~\in~ {\rm LLL}}
\phi^{(j)}  \phi^{(j)\dag} \otimes \mathbf{1}_z \otimes
\mathbf{1}_t =
\sum_{j~\in~ {\rm LLL}} \sum_{z, t} \psi^{(j)}_{z t}\psi^{(j)\dag}_{z t} \,.
  \end{aligned}
\end{equation}
In the last passage we have recast the
projector as a sum over modes localised on $x,y$ slices at fixed
$z,t$, namely $ \psi^{(j)}_{ z_0 t_0}(x,y,z,t) =
\phi^{(j)}(x,y)\delta_{z z_0}\delta_{t t_0}$. 
This construction carries over unchanged to the free lattice case,
provided the appropriate extra degeneracy of staggered fermions is
taken into account. 
Moreover, the formulation in terms of slice-localised states can be
exported with little modification to the interacting case. In fact, as
we have seen 
in the previous section, the LLL can be defined on 2D slices even in
the interacting case. If we now denote with $\psi^{(j)}_{z_0 t_0}$ the
solutions of the 2D Dirac equation on the $x,y$ slice at $z=z_0$ and
$t=t_0$ including the non-Abelian fields, the projector $P$
extracts precisely the contribution of the LLLs on all the slices to
the desired observable. Ordering the 2D modes according to their
eigenvalue, and taking into account the appropriate doubling
and colour factors we find for staggered fermions on the lattice in
the interacting case
\begin{equation}
  \label{eq:projqcd}
   P(B) = \sum_{j=1}^{2 N_c N_b}\sum_{z, t} \psi^{(j)}_{ z
  t}(B)\psi^{(j)\dag}_{ z t}(B)\,,
\end{equation}
where we have made explicit the dependence on the magnetic
field.\footnote{A similar projector for Wilson quarks in the free case
  is used in Ref.~\cite{Bignell:2017lnd}.} 
Our procedure consists essentially in changing basis from the usual
localised basis, where basis vectors are localised on the lattice
sites, to a new basis where the basis vectors are localised on an $x,y$
slice at fixed $z,t$, and solve the 2D Dirac equation on that slice, and
then extract the LLL contribution to observables by projecting on the
union of the modes corresponding to the 2D LLLs.

\begin{figure}[t]
  \centering
\subfigure{
  \includegraphics[width=0.43\textwidth,clip]{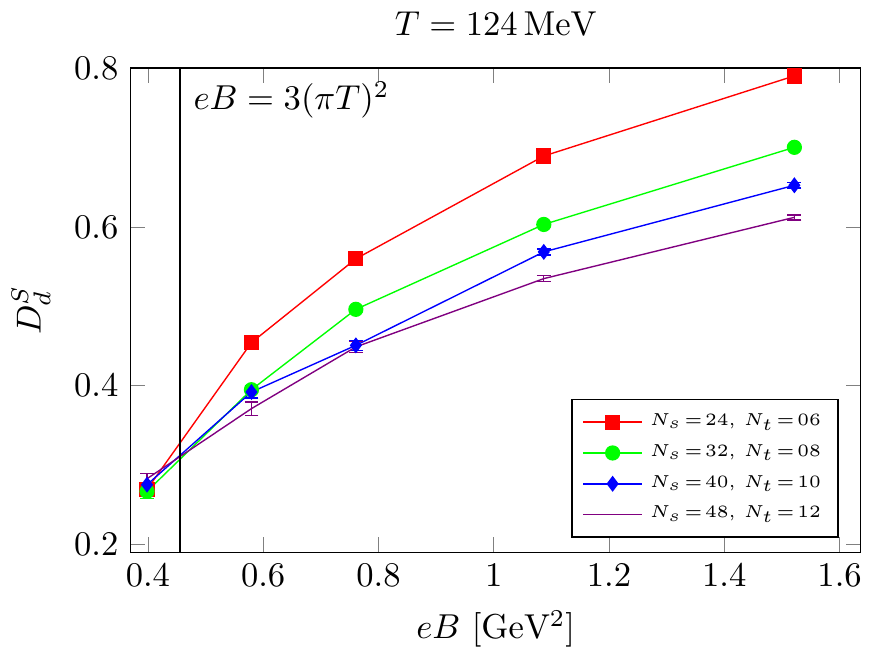}}
\subfigure{
  \includegraphics[width=0.43\textwidth,clip]{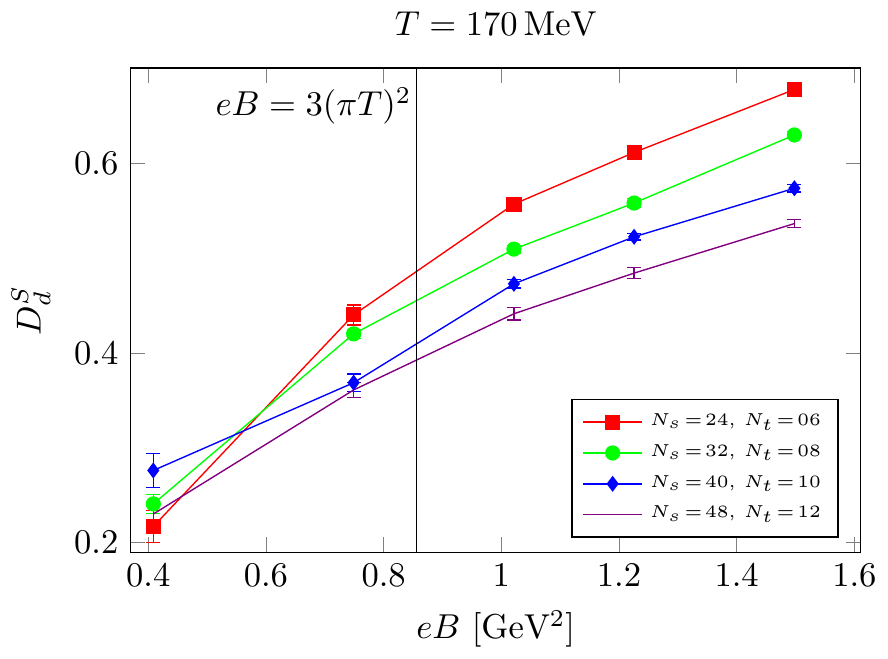}}
  \caption{The ratio $D^S_d$ of Eq.~\protect\eqref{eq:ratio} for the $d$
    quark as a function of the magnetic field at temperatures 
    below $T_c$ (left panel) and above $T_c$ (right panel).}
  \label{fig:DSfig}
\end{figure}

If the 2D LLLs have any physical effect we should be able to see some
distinctive feature in the overlap $\psi^{(i)\dag}_{z t}\phi$ between
the solutions $\phi$ of the 4D Dirac equation 
and the 2D LLL modes defined on the slices. 
Summing over slices and 2D doublers, we define the overlap factor as
\begin{equation}
  \label{eq:overlapdef}
W_i(\phi) = \sum_{\rm doublers} \sum_{z,t} 
\big|\psi^{(i)\dag}_{z t}\phi\big|^2\,.
\end{equation}
In Fig.~\ref{fig:overlap} we show $W_i(\phi)$ after averaging over a
small spectral interval of the 4D Dirac operator and over gauge
configurations. The overlap of low 4D  modes with the LLL modes is
almost independent of the 2D mode 
index, and jumps abruptly downwards at the end of the LLL; in
contrast, in the $B=0$ case this quantity is just continuous and
monotonically decreasing. For bulk modes the jump turns upwards, but
the flatness in the LLL remains. All modes turn out to have 
an LLL component, which is bigger for low modes. Notice that the
overlap with 2D modes seems to have a finite continuum limit.

\section{LLL approximation for the quark condensate}
\label{sec:cond}


Having defined the LLL in the 4D interacting case we can now return to
our main problem, and measure its contribution to the quark
condensate. We have already answered (mostly negatively) the
question whether the 
LL structure survives the introduction
of colour interactions, but since the LLL survives it is still
possible that it is indeed 
mostly responsible for MC. The LLL
contribution to the condensate is simply obtained by projecting the
Dirac operator on the LLL. Denoting with $\la\ldots\ra_B$ the 
average over gauge configurations in the presence of an external
magnetic field, the full and the projected condensates read
\begin{equation}
  \label{eq:cond_B}
  \la\bar\psi\psi\ra_B = \la \tr D^{-1}_{\rm stag}\ra_B\,,\qquad
\la\bar\psi\psi^{\rm  LLL}\ra_B =\la\tr\, P 
D_{\rm stag}^{-1}P \ra_B \,.
\end{equation}
However, 
both the full and the projected condensate are affected by UV
divergences, and they have to be renormalised.
For the full condensate
it is known that the change in the condensate due to a magnetic field, 
$\Delta\la\bar\psi\psi\ra(B)=\la\bar\psi\psi\ra_B -
\la\bar\psi\psi\ra_{B=0}$, is free from additive
divergences. On the other hand, subtracting the LLL projected 
condensate at $B=0$ from that at nonzero $B$ cannot cure the analogous
problem, since at $B=0$ the LLL is empty and $\la\bar\psi\psi^{\rm
  LLL}\ra_{B=0}=0 $ identically.  
Instead, we notice that from the 2D perspective the change in the full
condensate corresponds to the change in the contribution from all the
2D modes on all slices. Its analogue in the projected case is thus the
change in the contribution of the first $2 N_c N_b$ 2D modes on all
slices 
from zero to nonzero $B$, 
measuring the change in the condensate due to the change in the nature
of the modes which end up forming the LLL. Defining the new projector
\begin{equation}
  \label{eq:projqcd2}
   \widetilde{P}(B) = \sum_{j=1}^{2 N_c N_b}\sum_{z, t} \psi^{(j)}_{ z
  t}(0)\psi^{(j)\dag}_{ z t}(0)\,,
\end{equation}
which is $B$-dependent due to the upper limit in the sum, the subtracted
LLL-projected condensate is defined as
\begin{equation}
  \label{eq:cond_B_subproj}
 \Delta\la\bar\psi\psi^{\rm LLL}\ra(B) =
\la\bar\psi P(B)\psi\ra_B-\la\bar\psi\widetilde{P}(B)\psi\ra_0\,.
\end{equation}
It can be shown explicitly that this quantity is indeed free of
additive divergences in the free case. Since the additive divergence
comes from the large 4D modes where the SU$(3)$ interaction is negligible, we
expect that $ \Delta\la\bar\psi\psi^{\rm LLL}\ra(B)$ be free of
additive divergences also in the interacting case.

\begin{figure}[t]
  \centering
  \includegraphics[width=0.43\textwidth,clip]{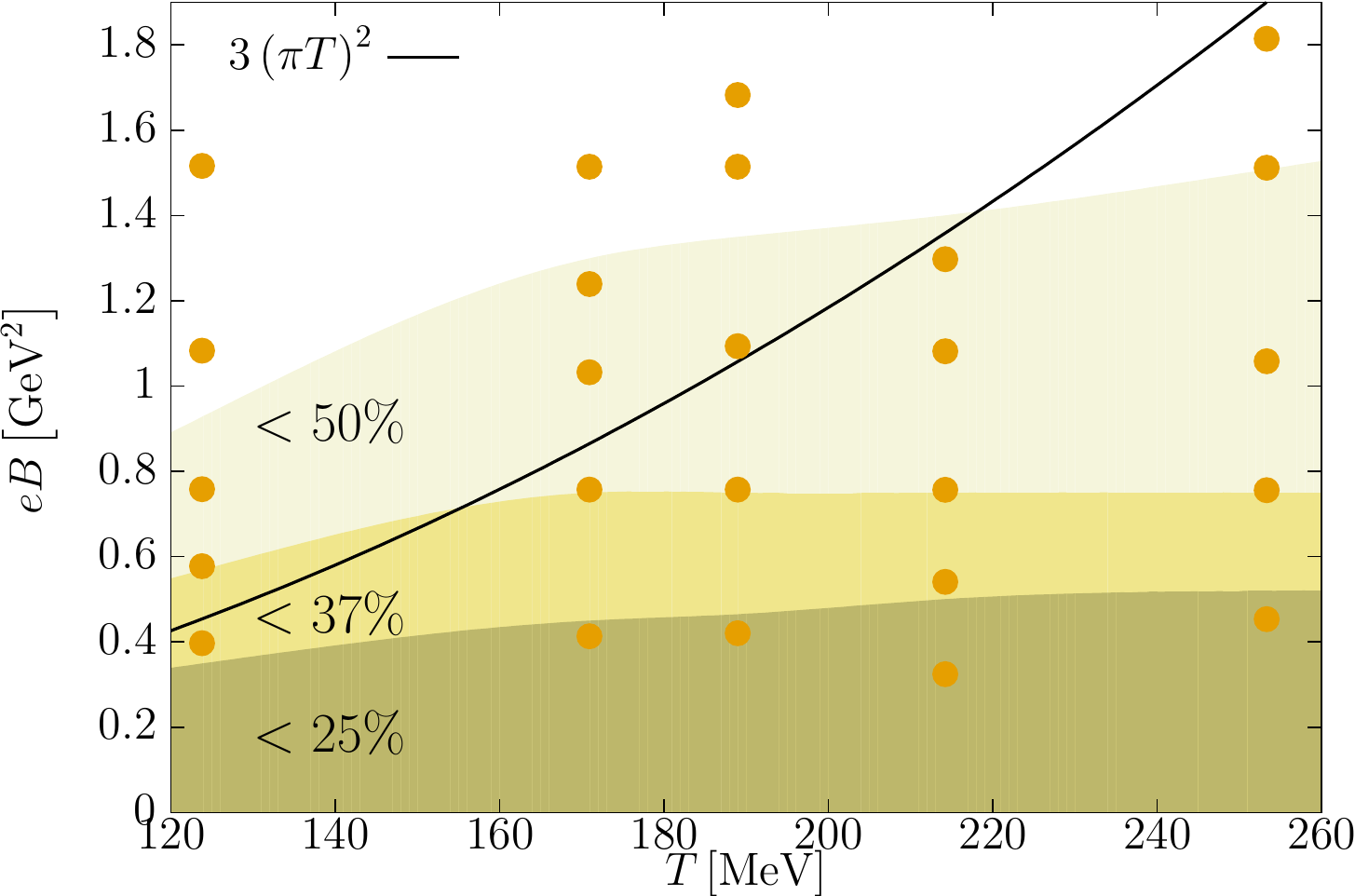}
  \caption{Visualisation of the validity of the LLL-approximation 
 for the $d$ quark condensate. The lighter the color, the closer the
 LLL-projected condensate is to the full result. The orange dots
 denote our simulation points and the  
 solid black line marks $|q_dB|=(\pi T)^2$.}
  \label{fig:summary}
\end{figure}

The multiplicative divergence can be studied by means of a
spectral decomposition of the projected condensate. If the overlap of
4D modes with the LLL, averaged over configurations and small
(infinitesimal) spectral intervals, has a continuum limit throughout
the spectrum both at zero and finite $B$, then the multiplicative
divergence is the same as in the full condensate, and therefore
cancels in ratios. The plots in Fig.~\ref{fig:overlap} indeed support
the existence of a continuum limit for the overlap. Nonetheless, such
a delicate issue requires more detailed studies than the present one,
and for this reason the continuum limit is not taken in this work.

Independently of the issue of renormalisability, the ratio of
the projected and full condensates after subtracting their additive
divergences,
\begin{equation}
  \label{eq:ratio}
   D^S =  \f{\Delta\la\bar\psi\psi^{\rm
      LLL}\ra(B)}{\Delta\la\bar\psi\psi\ra(B)} = 
\f{\la\bar\psi P(B)\psi\ra_B-\la\bar\psi\widetilde{P}(B)\psi\ra_0
}{\la\bar\psi\psi\ra_B-\la\bar\psi\psi\ra_0  } \,,
\end{equation}
measures how much of the change of the condensate is due
to the LLL at a given lattice spacing. If this ratio is
divergence-free, then it has full physical meaning, and it measures
the physical contribution of the LLL to the change of the condensate. 
By measuring it we can therefore answer our question about the role
of the LLL in the valence effect. If the LLL dominates the (change in
the) condensate, then this would naturally provide an explanation for
the valence effect (and thus for MC) as a consequence of the increased
degeneracy of the LLL. If the LLL does indeed dominate the condensate,
it is expected to do so on a 
configuration-by-configuration basis, and therefore it 
should not matter, at least from the qualitative point of view,
whether we include the magnetic field in the fermionic determinant or
not: this would just lead to a reweighting of the configurations. For
this reason, we decided to simplify our problem and consider only the
valence effect of the magnetic field, i.e., we included it in the
observables but not in the fermionic determinant. This should be
enough to decide whether the condensate is dominated by the LLL or
not. 



In Fig.~\ref{fig:DSfig} we show the results for $D^S$ at two
temperatures, one below and one above $T_c$, obtained for the
$d$ quark condensate ($q_d=-\f{e}{3}$) in our numerical simulations of
2+1 flavours of improved rooted staggered fermions 
(see
Ref.~\cite{Bali:2011qj} for technical 
details). 
The $B$ field was not included in the fermionic
determinant. 
We show results for various magnetic fields and different lattice 
spacings $a=1/(N_t T)$, where $N_t$ is the temporal extension of the
lattice in lattice units. The ratio $D^S_d$ seems to show a nice
scaling towards the continuum limit; increases with $B$, slowly
approaching 1, as one expects; and decreases with $T$. The results of
all our simulations are summarised in Fig.~\ref{fig:summary}, where we
show the regions 
of the $T-B$ plane where the change in the projected condensate
reaches a given fraction of the change in the full condensate. This is
estimated using our finest lattices, which, given the trend shown in
Fig.~\ref{fig:DSfig}, probably overestimates this fraction. This
allows to answer our question whether the LLL dominates or not the
change in the condensate: except for very large magnetic fields 
($eB\sim 1 \div 1.5 \,{\rm GeV}^2$) the LLL is responsible for
less than 50\% of the change in the condensate.

\section{Conclusions}
\label{sec:concl}

In this work we studied the issue of Landau levels on the lattice and
investigated the validity of the lowest-Landau-level
(LLL) approximation to QCD in the presence of background magnetic
fields. The presence of (nonperturbative) color interactions mixes the
levels, washing away the Landau level structure, with the exception of
the LLL. This can be defined in a consistent manner even
for strongly interacting quarks, thanks to a two-dimensional
topological argument that characterises the plane perpendicular to the
magnetic field. This allows to define a projector on the LLL that can
be used to extract its contribution to the observables. Observables in
4D are sensitive to the LLL, and in particular a sizeable fraction of
the change in the quark condensate comes from it. On the other hand,
the LLL does not dominate this change except for very large magnetic
fields, i.e., the LLL approximation typically underestimates the
change in the condensate. In other words, the LLL alone cannot fully
explain the valence effect and MC in most of the $T-B$ plane.

\bibliography{lllbib}

\end{document}